\documentclass[epjCONF,columns]{svjour}
\usepackage{graphics}
\usepackage[varg]{txfonts} 
\usepackage[latin1]{inputenc}
\session-title{Hadron Collider Physics Symposium 2011}

\begin{document}

\title{Standard Model Higgs Boson Combination at the Tevatron}

\author{Wei-Ming Yao\inst{1}\fnmsep\thanks{\email{wmyao@lbl.gov}}}
%
\institute{Physics MS-50B-5239, Lawrence Berkeley National Lab, 1 Cyclotron Rd,
 Berkeley, CA 94720, USA}

\abstract{
We present the recent results from combinations of searches for a 
standard model (SM) Higgs boson ($H$) by the CDF and D0 experiments at 
the Tevatron $p\bar p$ collider at $\sqrt{s}=1.96$~TeV. The data correspond to 
an integrated total luminosity of 8.2 (CDF) and 8.6 (D0) fb$^{-1}$. Compared to
the previous Tevatron Higgs boson search combination more data have been added,
additional channels have been added, and some previously used channels have 
been reanalyzed to gain sensitivity.   
No excess is observed above background expectation, and set 95\% C.L. upper 
limits (median expected) on Higgs boson production at factors of 1.17 (1.16), 
1.71 (1.16), and 0.48(0.57) times the SM predictions for Higgs bosons of 
mass $m_H$=115, 140, and 165 GeV/c$^2$, respectively. We exclude a 
standard-model Higgs boson in the mass range 156--177 GeV/c$^2$ at the 
95\% C.L, with an expected exclusion region 148--180 GeV/c$^2$. 
The absence of a Higgs boson signal also constrains some new physics such 
as $4^{th}$ generation models and other exotic models. 
} 
\maketitle
\section{Introduction}
\label{intro}
The Higgs boson is the last unobserved particle postulated in SM to 
help explain the origin of mass. The global fit from the electroweak 
precision data indicates that the Higgs boson mass is lighter than 158 GeV/c$^2$ at 
95\% confidence level~\cite{ewk}. The direct searches from LEP~\cite{lep}, the 
Tevatron~\cite{tev11} and LHC~\cite{lhc} further limit the Higgs boson 
mass between 114.4 and 141 GeV/c$^2$. Recently, using 5 fb$^{-1}$ of integrated 
luminosity collected in 2011 for each experiment, the ATLAS and CMS collaborations~\cite{excess} have 
reported excesses in the Higgs search channels ($\gamma\gamma, ZZ^*\rightarrow 4l$, and $WW^*\rightarrow 2l$)
with reconstructed invariant mass near 125 GeV/c$^2$. The combined statistical significance is about 2.0 $\sigma$ 
including the look elsewhere effect, but nevertheless, this is interesting and requiring more data to confirm or 
reject the hypothesis. 

With full dataset and 
improved analyses, The Tevatron will be still competitive and will provide an 
unique sensitivity to $H\rightarrow b\bar b$ in the remaining mass range.   
In this note, we present the recent results from the combination of multiple 
direct searches for the SM Higgs boson at the Tevatron~\cite{tev11}. 
The analyses that are combined seek signals of Higgs bosons produced in 
association with vector bosons ($q\bar q\rightarrow W/Z H$), through 
gluon-gluon fusion ($gg\rightarrow H$), and through vector boson fusion 
(VBF) ($q\bar q\rightarrow q'\bar q' H$) corresponding to
integrated luminosities up to 8.2 fb$^{-1}$ at CDF 
and 8.6 fb$^{-1}$ at D0.   

\section{Higgs Boson Search Strategies}
\label{sec:1}
 The dominant Higgs boson production processes at the Tevatron are 
 gluon-gluon fusion, and associated production with a $W$ or $Z$ boson. For 
 a mass below 135 GeV/c$^2$, the Higgs boson decays predominantly 
 into $b\bar b$, 
 which makes the associated production most assessible while the 
direct production of $gg\rightarrow H \rightarrow b\bar b$ is 
 limited by multi-jet QCD background. For a mass above 135 GeV/c$^2$, 
 the Higgs boson will decay predominately into $WW^*$ and $ZZ$, making the 
 $gg\rightarrow H$ production most useful. However, the best sensitivity 
 requires combining all production and decay channels together, 
 including both CDF and D0 data.  

  The search strategies for the Higgs boson are quite similar for the 
  corresponding CDF and D0 analyses. For low mass signatures we look for 
  a $b\bar b$ mass resonance in associated with $W/Z$ events where $W/Z$ decays
  leptonically. We apply b-tagging and advanced multivariate analysis (MVA) 
  technique to suppress large $W/Z$ +jets and top background. 
  For high mass signatures we look for the Higgs boson decaying into 
  $WW$ pair in the inclusive Higgs boson events that lead to many interesting 
  final states. The most sensitive channel is both $W$ decaying leptonically 
  that gives a final state of opposite-sign dilepton, large missing Et, and 
  some jets. Due to missing neutrinos, we have to rely on the event kinematic 
  distributions that distinguish the signal from the background 
  using MVA techniques. There are 
  in total 165 mutually exclusive final states, 94 channels from D0 and 71 
  channels from CDF.

  All analyses provide binned histograms of the final discriminant for data,
  signal, and each individual background. More details for the low and 
  high-mass SM Higgs boson searches can be found in these talks~\cite{others}. 

  We use the most recent high-order calculations of the SM Higgs boson 
  production cross section and decay branching ratio to normalize the 
  signal event yield in each individual channel. So we can combine them 
  statistically.

\subsection{Combination Procedures}
\label{sec:1.1}
To gain confidence that the final result does not depend on the details of 
statistical  method, we perform two types of combinations, using Bayesian and 
modified Frequentist approaches, which yield results that 
agree within 10\%. We quote only the limits obtained with the Bayesian method, which is
decided upon a priori. 

Both methods rely on distributions of final discriminants, 
not just on event counts. 
Systematic uncertainties are treated as nuisance parameters with truncated 
Gaussian. Both methods use likelihood calculations based on Poisson 
probabilities. There are two types of systematic uncertainties that affect 
the rate and shape of estimated  signal and background in a correlated way. 
The rate systematic only affects overall normalization while the shape 
systematic is changing differential distribution due to the jet energy 
scale(JES) and Monte Carlo(MC) modeling.

CDF and D0 share common systematic uncertainties on luminosity, the 
theoretical cross sections, and some scale and PDF veriations, which are 
treated as correlated. 
Other sources of systematic are experiment dependent, treated uncorrelated 
between experiments, but correlated within the experiment, such as 
lepton identification, b-tagging efficiency, JES, detector effects and 
instrumental backgrounds.

\begin{figure}
\begin{center}
\resizebox{1.0\columnwidth}{!}{%
  \includegraphics{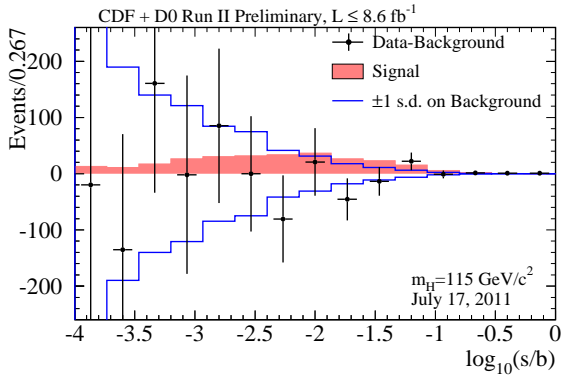} 
}
\caption{Background-subtracted distributions of $log_{10}(s/b)$ for the data from all contributing channels from the combined 
CDF and D0 analyses for $m_H=115$ GeV/c$^2$.}
\label{fig:sb115}       
\end{center}
\end{figure}

\begin{figure}
\begin{center}
\resizebox{1.0\columnwidth}{!}{%
   \includegraphics{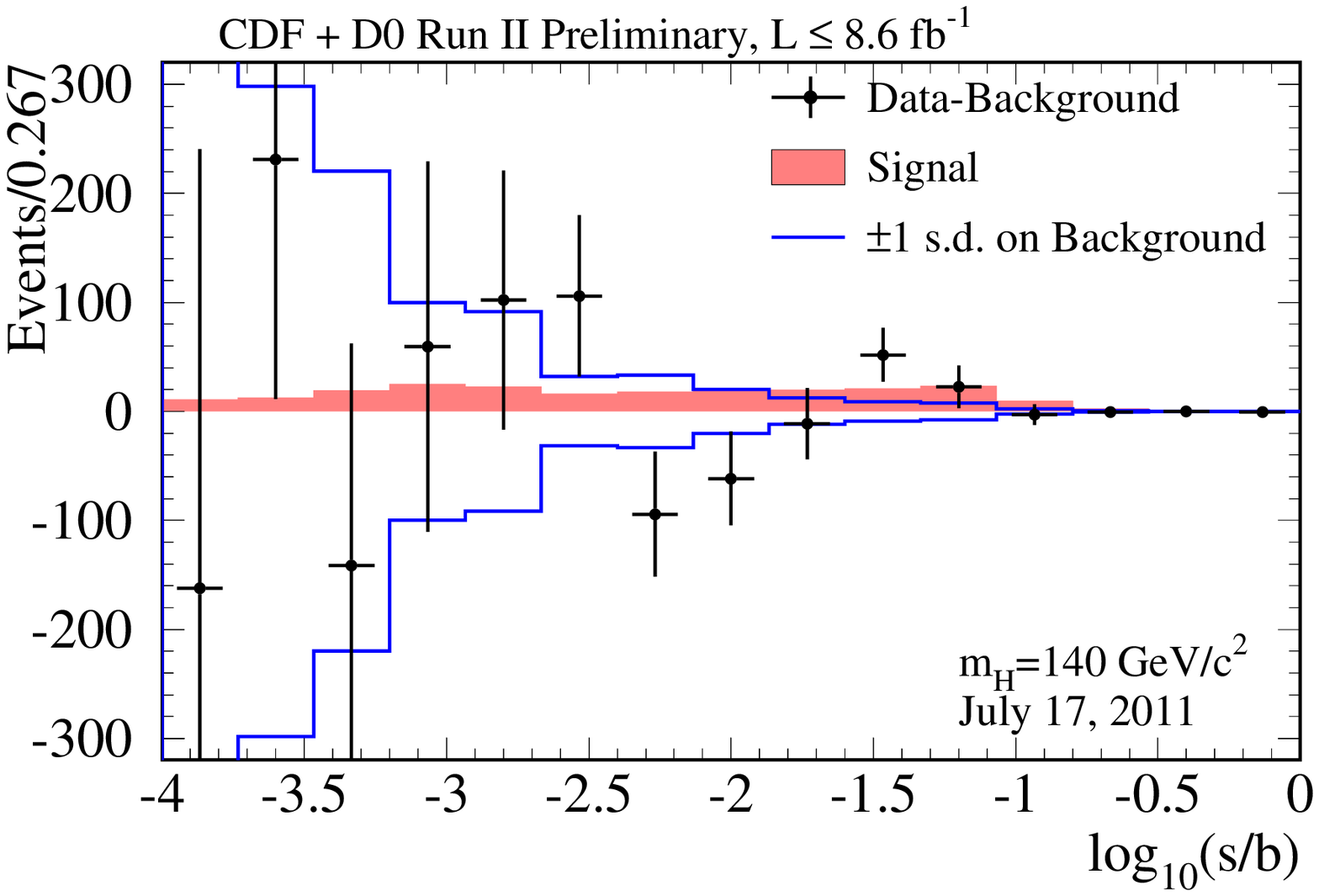} 
}
\caption{Background-subtracted distributions of $log_{10}(s/b)$ for the data from all contributing channels from the 
combined CDF and D0 analyses for $m_H=140$ GeV/c$^2$.}
\label{fig:sb140}       
\end{center}
\end{figure}

\begin{figure}
\begin{center}
\resizebox{1.0\columnwidth}{!}{%
  \includegraphics{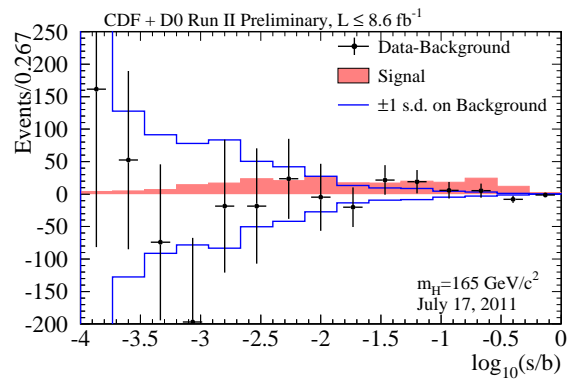} 
}
\caption{Background-subtracted distributions of $log_{10}(s/b)$ for the data from all contributing channels from the 
combined CDF and D0 analyses for $m_H=165$ GeV/c$^2$.}
\label{fig:sb165}       
\end{center}
\end{figure}

In order to check the consistency between data and expectations, 
we rebin the final discriminant from each channel in terms of signal to 
background ratio (s/b), data with similar s/b may be added without 
loss in sensitivity. Figure~\ref{fig:sb115}~-~\ref{fig:sb165} show the data after the background 
subtraction, compared to the expected signal as function of log(s/b) for 
$m_H =115$, 140, and 165 GeV/$c^2$, respectively.  There are no significant 
excess of events observed in any of the higest s/b bins.

\section{Combined Tevatron Searches for the SM Higgs Boson} 
\label{sec:2} 
\begin{figure}
\begin{center}
\resizebox{1.0\columnwidth}{!}{%
  \includegraphics{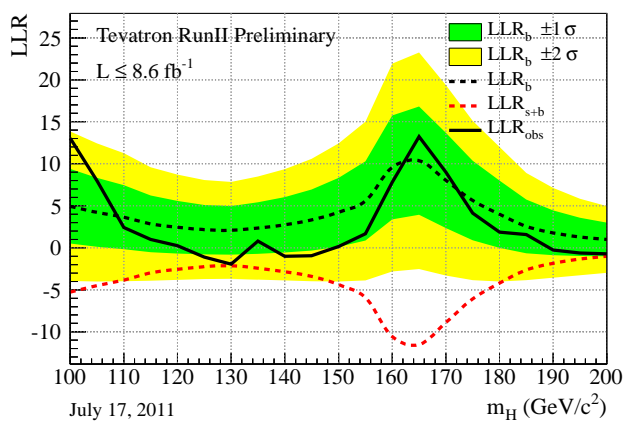} 
}
\caption{Distributions of the log-likelihood ratio as a function of Higgs boson mass
obtained with the CL$_s$ method for the Tevatron combination.
The bands indicate the region enclosing 1-$\sigma$ and 2-$\sigma$ fluctuations of the background, respectively.}
\label{fig:limita}       
\end{center}
\end{figure}

Before extracting the combined results, we check the search sensitivity 
using log-likelihood ratio(LLR) for different hypotheses to quantify the 
expected sensitivity across the mass range tested.  
Figure~\ref{fig:limita} shows the combined distributions of the 
log-likelihood ratio as a function of Higgs boson mass.  
The black dot curve is for the 
background-only hypothesis, the red dot curve is for the signal-plus-background 
hypothesis, and the solid curve is for the observed data. The sizes of one 
and two sigma bands indicate the width of the LLR background-only 
distribution. The separation between the background-only and 
signal-plus-background hypotheses provides a measure of the search 
sensitivity, which is about two sigma at low mass and close to  
3 sigma at $m_H=165$ GeV/c$^2$.

\begin{figure}
\begin{center}
\resizebox{1.0\columnwidth}{!}{%
  \includegraphics{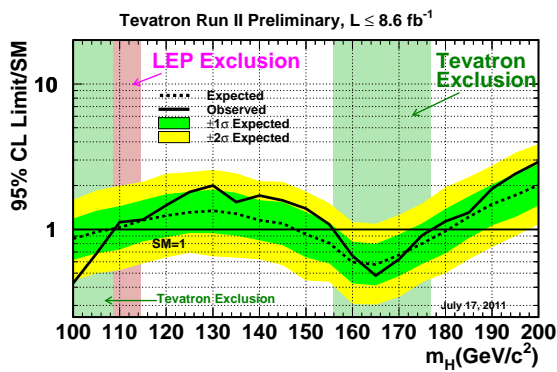} 
}
\caption{Observed and expected 95\% C.L. upper limits on the ratio to the SM prediction, 
as a function of the Higgs boson mass for the combined CDF and D0 analyses. 
The bands indicate the 68\% and 95\% probability regions where the limits 
can fluctuate, in the absence of signal. }
\label{fig:limitb}       
\end{center}
\end{figure}

Figure~\ref{fig:limitb} shows the ratio of the 95\% C.L. expected 
and observed limit to the SM Higgs boson cross section times branching ratio at 
the Tevatron after combining CDF and D0 searches together~\cite{tev11}.
The observed and median expected ratios are listed in Table~\ref{tab:low} for 
$m_H\le 150$ GeV/c$^2$, and in Table~\ref{tab:hi} for $m_H\ge 155$ GeV/c$^2$, 
as obtained by the Bayesian and the CL$_s$ methods.  
We obtain the observed limit of 1.17 with 
expected 1.16 for $m_H=115$ GeV/c$^2$ and 0.48 with expected 0.57 
for $m_H=165$ GeV/c$^2$.

\begin{figure}
\begin{center}
\resizebox{1.0\columnwidth}{!}{%
  \includegraphics{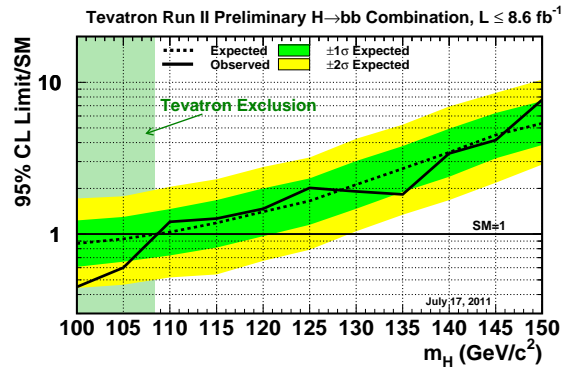} 
}
\caption{Observed and expected 95\% C.L. upper limits on the ratio to the SM prediction for the $H\rightarrow b\bar b$ decay. 
The bands indicate the 68\% and 95\% probability regions where the limits 
can fluctuate, in the absence of signal. }
\label{fig:limitx}       
\end{center}
\end{figure}

We investigate the sensitivity and observed limits using CDF's and D0's 
searches for $H\rightarrow b\bar b$ taken in combination. These channels 
contribute the most for values of $m_H$ below 135 geV/c$^2$ and will remain 
competitive with the LHC experiments. The result of this combination is 
shown in Figure~\ref{fig:limitx}.

\begin{table}
\caption{Ratios of median expected and observed 95\% C.L. limit to the 
SM cross section for the combined CDF and D0 analyses as a function of the Higgs boson mass.}
\label{tab:low}       
\begin{tabular}{llllllll}
\hline\noalign{\smallskip}
Bayesian & 100 & 110 & 115 & 120 & 130 & 140 & 150  \\
\noalign{\smallskip}\hline\noalign{\smallskip}
Expected & 0.86 & 1.03 & 1.16 & 1.24 & 1.35 & 1.16 & 0.93 \\
Observed & 0.43 & 0.68 & 1.12 & 1.47 & 2.00 & 1.71 & 1.39 \\
\noalign{\smallskip}\hline
\end{tabular}
\end{table}

\begin{table}
\caption{Ratios of median expected and observed 95\% C.L. limit to the 
SM cross section for the combined CDF and D0 analyses as a function of the Higgs boson mass.}
\label{tab:hi}       
\begin{tabular}{llllllll}
\hline\noalign{\smallskip}
Bayesian & 155 & 160 & 165 & 170 & 180 & 190 & 200  \\
\noalign{\smallskip}\hline\noalign{\smallskip}
Expected & 0.80 & 0.59 & 0.57 & 0.67 & 0.97 & 1.49 & 2.02 \\
Observed & 1.08 & 0.66 & 0.48 & 0.62 & 1.14 & 1.90 & 2.91 \\
\noalign{\smallskip}\hline
\end{tabular}
\end{table}

\section{Constraints on Fourth-Generation and other Exotic Models}
\label{sec:3}

\begin{figure}
\begin{center}
\resizebox{1.0\columnwidth}{!}{%
  \includegraphics{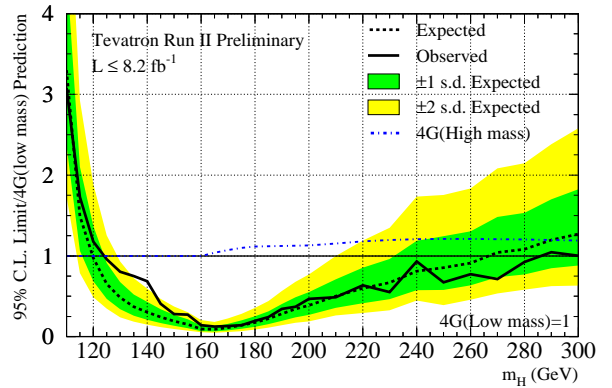} 
}
\caption{The combined observed (solid black lines) and expected 
(dashed black lines) 95\% C.L. upper limit on 
$\sigma(p\bar p\rightarrow H)\times B(H\rightarrow W^+W^-)$ are shown in the 
4th generation model as a function of the Higgs boson mass.}
\label{fig:4g}       
\end{center}
\end{figure}

\begin{figure}
\begin{center}
\resizebox{1.0\columnwidth}{!}{%
  \includegraphics{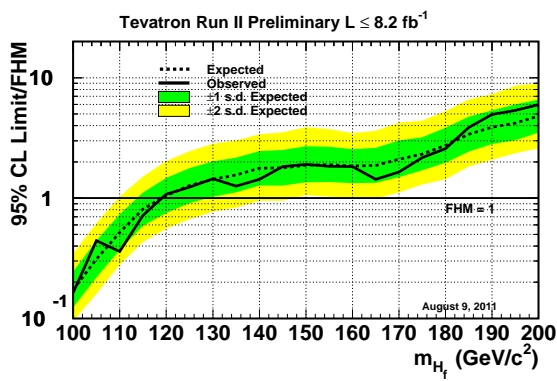} 
}
\caption{The combined observed (solid black lines) and expected 
(dashed black lines) 95\% C.L. upper limit on 
$\sigma(p\bar p\rightarrow H)\times B(H\rightarrow W^+W^-)$ are shown
in a fermiphobic model as a function of the Higgs boson mass.}
\label{fig:gg}       
\end{center}
\end{figure}

With the absence of a $H\rightarrow WW$ signal, we can use it to set constraints 
on 4th generation and other exotic models~\cite{4g}~\cite{fermiophobic}. 
For example, in a 4th generation 
model, the Higgs boson production cross section of $gg\rightarrow H$ could be 
enhanced by a factor of 9 due to additional 2 heavy quarks contributions 
in the loop. Reinterpreting $H\rightarrow WW$ limit in the 
4th generation model, as shown in Figure~\ref{fig:4g} as a function of
the Higgs boson mass, we are able to exclude the Higgs boson mass between 124 
and 286 GeV/c$^2$ at 95\% CL,  Further combining $H\rightarrow \gamma\gamma$, 
$WW$ limit in a fermiphobic model, as shown in Figure~\ref{fig:gg} 
as a function of the Higgs boson mass, we are able to exclude the Higgs boson 
mass above 119 GeV/c$^2$ at 95\% CL. 

\section{Conclusion and Future Prospects}
\label{sec:4}

We have presented the recent combinations of searches for a 
standard model (SM) Higgs boson ($H$) by the CDF and D0 experiments at 
the Tevatron $p\bar p$ collider at $\sqrt{s}=1.96$~TeV. With up to 8.2 fb$^{-1}$
of data analyzed at CDF and up to 8.6 fb$^{-1}$ at D0, the 95\% C.L. upper
limits (median expected) on Higgs boson production are factors of 1.17 (1.16), 
1.71 (1.16), and 0.48(0.57) times the SM predictions for Higgs bosons of 
mass $m_H$=115, 140, and 165 GeV/c$^2$, respectively. We exclude a 
standard-model Higgs boson in the mass range 156--177 GeV/c$^2$ at the 
95\% C.L, with an expected exclusion region 148--180 GeV.c$^2$. 
The absence of Higgs boson signal also constrains some new physics such 
as $4^{th}$ generation models and other exotic models as well.  

The Tevatron has ended on 9/30/2011 for the last time after 28 years 
With 10 fb-1 analyzable dataset and anticipated improvement, the Tevatron will
remain competitive to reach 95\% C.L. exclusion sensitivity over the Higgs 
boson mass range up to 185 GeV/c$^2$. In addition, we combine the CDF and 
D0 analyses which seek specifically the $H\rightarrow b\bar b$ decay, 
which dominates at the low end of the allowed mass range for the SM 
Higgs boson and will remain competitive with the LHC experiments for 
several years to come.

\section*{Acknowledgments}
\label{sec:5}
We would like to thank the organizers of the 2011 Hadron Collider Physics 
Symposium for a wonderful conference with excellent presentations and the CDF and D0 collaborations 
for the results presented at this conference.


\begin{thebibliography}{99}
%

\bibitem{ewk} 
LEP-Tevatron-SLD Electroweak Working Group, arXiv:0911.2604. 

\bibitem{lep}
R. Barate et al (LEP Higgs Working Group), Phys. Lett. B \textbf{565}, 
(2003) 61.

\bibitem{tev11}
The CDF and D0 Collaborations and the TEVNPHWG Working Group, 
arXiv:1107.5518 (2011). 

\bibitem{lhc}
G. Rolandi, ``Higgs Status (Tevatron+LHC) and combination'', this proceeding;\\
ATLAS Collaboration, ATLAS-CONF-2011-157;\\
CMS Collaboration CMS-PAS-HIG-11-023.

\bibitem{excess}
F. Gianotti (ATLAS Collaboration) and G. Tonelli (CMS Collaboration), 
Talk at CERN Public Seminar on Dec 13, 2011.    


\bibitem{others} 
F. Sforza, ``Searches for Low-Mass SM Higgs at the 
Tevatron'', this proceeding; \\
K. Petridis, ``Searches for High-Mass SM Higgs at 
the Tevatron'', this proceeding. 

\bibitem{4g}
The CDF and D0 Collaborations and the TEVNPHWG Working Group, 
arXiv:1108.3331 (2011).
   
\bibitem{fermiophobic}
The CDF and D0 Collaborations and the TEVNPHWG Working Group, 
FERMILAB-CONF-11-413-E (2011). 

\end{thebibliography}
\end{document}